\documentclass[fleqn,12pt,twoside]{article}
\usepackage{espcrc1}
\usepackage{here}
\usepackage{wrapfig}

\usepackage{graphicx}

\newcommand{\AmS}{{\protect\the\textfont2
  A\kern-.1667em\lower.5ex\hbox{M}\kern-.125emS}}

\title{A search for deeply bound kaonic nuclear states }
\author{T. Suzuki\address[UT]{Department of Physics, University of Tokyo, Hongo, Bunkyo-ku, Tokyo 113-0033, Japan},
       H. Bhang\address[SNU]{Department of Physics, Seoul National University, Shikkim-dong, Kwanak-gu, Seoul 151-742, Korea}, G. Franklin\address[CMU]{Department of Physics, Carnegie Mellon University, Pittsburgh, PA 15213, USA}, K. Gomikawa\addressmark[UT], R.S. Hayano\addressmark[UT],\\
T. Hayashi\address[TIT]{Department of Physics, Tokyo Institute of Technology, Ookayama, Meguro-ku, Tokyo 152-8551, Japan}\thanks{Present address: Department of Legal Medicine, Osaka University}, K. Ishikawa\addressmark[TIT], S. Ishimoto\address[KEK]{IPNS, KEK(High Energy Acceralator Research Organization), Oho, Tsukuba-shi, Ibaraki 305-0801, Japan}, K. Itahashi\address[RIKEN]{DRI, RIKEN, Wako-shi, Saitama 351-0198, Japan}, M. Iwasaki\addressmark[RIKEN]\addressmark[TIT], T. Katayama\addressmark[TIT],\\
Y. Kondo\addressmark[TIT], Y. Matsuda\addressmark[RIKEN], T. Nakamura\addressmark[TIT], S. Okada\addressmark[TIT]\thanks{Present address: DRI, RIKEN}, H. Outa\addressmark[KEK]\thanks{Present address: DRI, RIKEN}, B. Quinn\addressmark[CMU],\\
M. Sato\addressmark[TIT], M. Shindo\addressmark[UT], H. So\addressmark[SNU], P. Strasser\addressmark[RIKEN]\thanks{Present address: IMSS, KEK}, T. Sugimoto\addressmark[TIT], K. Suzuki\addressmark[UT]\thanks{Present address: Physik Department E12, Technishe Universit\"at M\"unchen}, S. Suzuki\addressmark[KEK],\\
D. Tomono\addressmark[TIT]\thanks{Present address: IMSS, KEK}, A.M. Vinodkumar\addressmark[TIT], E. Widmann\addressmark[UT]\thanks{Present address: Stefan Meyer Instituite for Subatomic Physics, Wien}, T. Yamazaki\addressmark[RIKEN], and T. Yoneyama\addressmark[TIT] 
}
\begin{document}

\maketitle

\begin{abstract}
We have measured nucleon energy spectra by means of time-of-flight (TOF) from ${}^4$He($K^-_{stopped},\thinspace N$) reactions (KEK PS E471 experiment). In the proton spectrum, a clear mono-energetic peak was observed under semi-inclusive condition, which was assigned to the formation of a strange tribaryon S${}^0$(3115) with isospin $T=1$. The mass and width of the state were deduced to be $3117.7^{+3.8}_{-2.0}(syst) \pm 0.9(stat)$ MeV/$c^2$ and $<$ 21.6 MeV/$c^2$, respectively, and its main decay mode was $\Sigma NN$. In the neutron spectrum, a mono-energetic peak was found as the result of a detailed analysis, which was assigned to the formation of another kind of strange tribaryon S${}^+$(3140). The mass and width of the state were deduced to be $3140.5^{+3.0}_{-0.8}(syst) \pm 2.3(stat)$ MeV/$c^2$ and $<$ 21.6 MeV/$c^2$, respectively, and its main decay mode was $\Sigma^{\pm} NN$. The isospin of the state is assigned to be 0. The results are compared with recent theoretical calculations.

\end{abstract}
\vspace{5mm}

\section{Introduction}
Recentry, we have discovered two kinds of {\it strange tribaryons} by measuring nucleon energy spectra from the stopped $K^-$ reaction on ${}^4$He \cite{E471NIM}, which was motivated by the prediction of a deeply bound $K^-ppn$ state by Akaishi and Yamazaki \cite{Akaishi-Yamazaki}. The first kind, S${}^0$(3115), was discovered in
\begin{equation}
(K^- - {}^4{\rm He})_{atomic} \rightarrow {\rm S}^0(3115) + p
\end{equation}
reaction \cite{E471PLB_P}. The observed state has isospin $T=1$, charge $Z=0$ and mass $M_{{\rm S}^0} \sim 3118$ MeV/$c^2$.
The second kind, S${}^+$(3140), was indicated from
\begin{equation}
(K^- - {}^4{\rm He})_{atomic} \rightarrow {\rm S}^+(3140) + n
\end{equation}
reaction \cite{E471PLB_N} originally proposed to search for the $K^-ppn$ state, which was predicted to be at $M$ $=$ $3194$ MeV/$c^2$ with $T=0$ and $Z=1$ \cite{Akaishi-Yamazaki}.

In this paper, we summarize the experimental results and compare them to some theories.

\section{Experimental principle}
 We adopted the stopped $K^-$ method on ${}^4$He to search for the predicted $K^-ppn^{Z=1}_{T=0}$ state \cite{E471NIM,Akaishi-Yamazaki}. A stopped $K^-$ forms a kaonic atom and is absorbed by the nucleus after an electromagnetic cascade. If it, or more generally, strange tribaryonic states with $Z=1$ which are denoted S${}^+$, exist, they will be formed via
\begin{equation}\label{eqn1}
 (K^- - {}^4{\rm He})_{atomic} \rightarrow {\rm S}^+ + n,
\end{equation}
and appear as distinct peaks in the emitted neutron energy spectrum. By this reaction, both $T=0$ and $T=1$ tribaryonic states can be populated. On the otherhand, by similar reaction,   
\begin{equation}\label{eqn2}
 (K^- - {}^4{\rm He})_{atomic} \rightarrow {\rm S}^0 + p,
\end{equation}
$T=1$ neutral tribaryonic states, S${}^0$, can be exclusively populated. Those strange tribaryonic states are generally expected to decay strongly into $\Lambda \pi NN$, $\Sigma NN$, or $\Lambda NN$, and secondary charged particle $\pi^{\pm}$ or $p$ should appear in the final state. We measured the nucleon energy spectra by means of time of flight (TOF), in coincidence with incident kaons and secondary charged particles from the reaction. Hence, it should be noted that measured spectra are {\it semi-inclusive} ones in that meaning. A detailed description of the experimental setup is given in Refs \cite{E471NIM,E471PLB_P,E471PLB_N}. 

\section{${}^4$He($K^-_{stopped},p$) missing mass spectrum and S${}^0$(3115)}
\begin{wrapfigure}[25]{l}{10cm}
     \includegraphics[width=9cm]{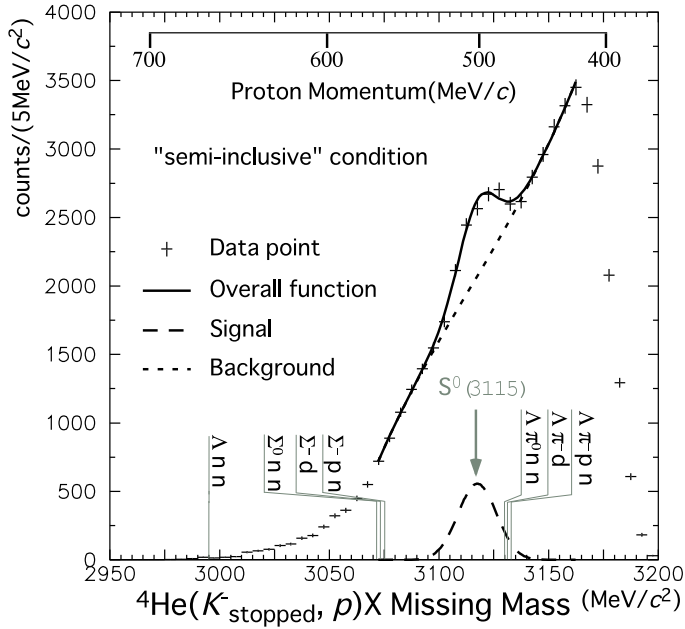}
  \caption{Missing mass spectrum of the ${}^4$He($K^-_{stopped},\thinspace p$) reaction under semi-inclusive condition \cite{E471PLB_P}. Resulting fit functions are shown by solid, dotted and dashed lines to represent all, background and signal parts, respectively.}\label{p_missingmass}
\end{wrapfigure}

A ${}^4$He($K^-_{stopped},\thinspace p$) missing mass spectrum under semi-inclusive condition is given in Fig. \ref{p_missingmass}. The details of the analysis - identification of charged particle on neutron counters, TOF analysis, {\it etc}, are given in Ref. \cite{E471PLB_P} and a future publication. It exhibits a clear peak structure at around 3118 MeV/$c^2$ on a broad continuum background which mainly originates from non-mesonic absorption process ($p_{nm}$) or subsequent $\Lambda$ decay ($p_{nm-\Lambda D}$), $K^- \tilde{N} \tilde{N} \rightarrow  p_{nm} \thinspace Y_{nm}$, and $\Lambda_{nm}  \rightarrow  p_{nm-\Lambda D} \thinspace \pi^-$, where $\tilde{N}$ and $Y$ denote a bound nucleon and hyperon($\Lambda, \Sigma$), respectively. From the detailed analysis of the spectrum, it is clearly shown that this peak structure\footnote{Saying more strictly, at least a certain large fraction of the peak counts.} is not due to hypernuclear formation and its two-body decay, $(K^- + {}^4{\rm He})_{atomic} \rightarrow \pi^- + {}^4_{\Lambda}{\rm He}$, ${}^4_{\Lambda}{\rm He} \rightarrow p + {}^3{\rm H}$
and the peak is assigned to the formation of a neutral strange tribaryonic state S${}^0$ with $T=1$, which has never been observed \cite{E471PLB_P}.

 In order to determine the mass and width of the state, we fitted a Gaussian peak plus a third-order polynomial background to the 19 data points in the peak region as indicated in the figure. The fit $\chi^2$ is 8.76 for 12 degrees of freedom and the statistical significance of the peak, defined as the signal area divided by its fit error, is 8.2 $\sigma$. The fitted mass is 3117.7 $\pm$ 0.9(stat) MeV/$c^2$, while our estimate for the systematic uncertainty in the absolute mass is $^{+3.8}_{-2.0}$ MeV/$c^2$, and the observed state is denoted S${}^0$(3115), expressing the mass in the name. The Gaussian standard deviation in the fitting is 8.7 $\pm 0.7 (stat) {}^{+1.3}_{-0.3} (syst)$ MeV/{\it c}$^2$.  The upper- and lower-limit of TOF resolution, 300 and 200 psec, respectively, correspond to a mass resolution of 8.6 and 6.4 MeV/$c^2$ ($\sigma$) in the region of interest. Using the lower limit of the instrumental resolution, the natural width $\Gamma_{\rm S^0}$ is evaluated to be smaller than 21.6 MeV/$c^2$ at 95 $\%$ confidence level. The main decay mode of the state is deduced to be $\Sigma NN$ from the study of secondary charged particle detected in coincidence with the signal, and its formation probability per stopped $K^-$ is estimated to be several$\times 10^{-3}$. The branching ratios of the decay and the formation will be discussed in a forthcoming paper.

\section{${}^4$He($K^-_{stopped},\thinspace n$) missing mass spectrum and S${}^+$(3140)}
\begin{wrapfigure}[40]{l}{9cm}
  \begin{center}
     \includegraphics[width=9cm]{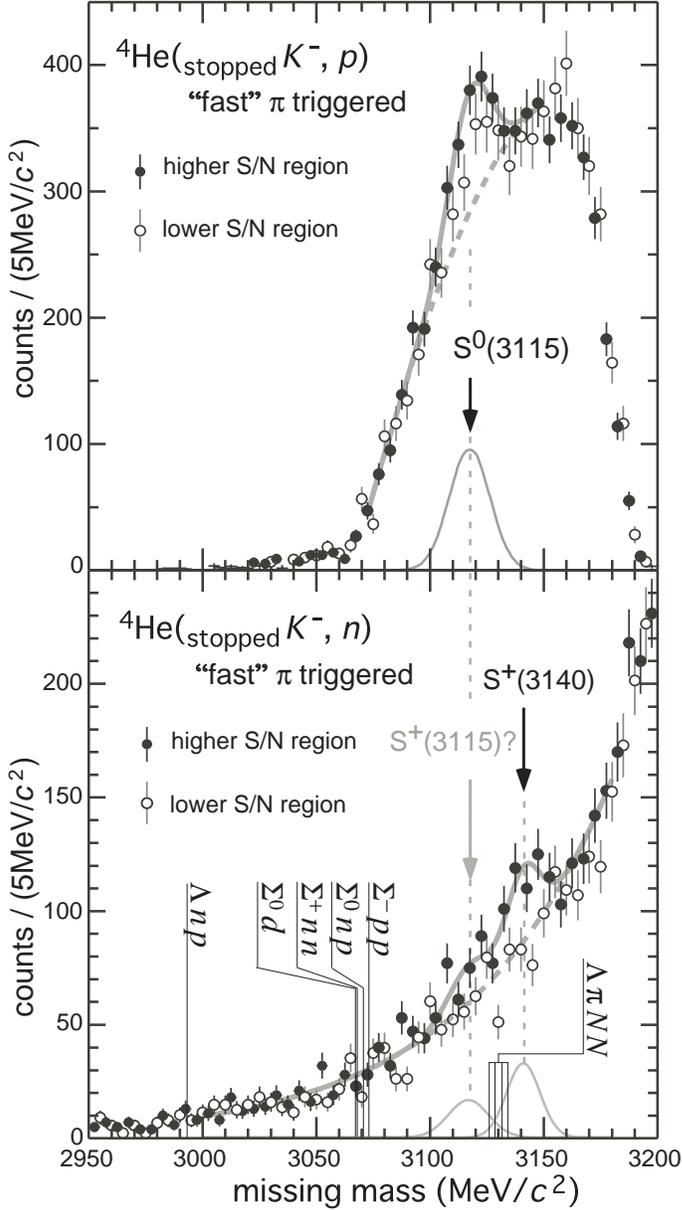}
  \end{center}
  \caption{Missing mass spectra of the ${}^4$He($K^-_{stopped},p$) (top) and ${}^4$He($K^-_{stopped},n$) (bottom) reactions \cite{E471PLB_N}.}\label{n_missingmass}
\end{wrapfigure}

The measured neutron spectrum can be analyzed in connection with the proton spectrum. The ${}^4$He($K^-_{stopped},\thinspace p$) and ${}^4$He($K^-_{stopped},\thinspace n$) spectra under higher-S/N and lower-S/N conditions are given in the top and bottom of Fig. \ref{n_missingmass}, respectively.  A comprehensive description of higher- and lower-S/N conditions, which are the event selection criteria related to the vertex inconsistency and secondary charged particle momentum to achieve higher- and lower-S/N ratio, is given in Ref. \cite{E471PLB_N}. 

In comaprison of both the top and bottom, the total statistics between higher- and lower-S/N data points are set identically. On the top, we see a clear enhancement of the S${}^0$(3115) component in the comparison between higher- and lower-S/N data points, hence the proper work of the selection criteria is confirmed. On the bottom, an enhancement is observed at around 3140 MeV/$c^2$ on a substantial continuum, which is expected to mainly originate from non-mesonic absorption process, and the structure indicates the formation of another kind of strange tribaryon, S${}^+$(3140). In this case, no other processes can generate neutrons with similar momentum, so that the interpretation is somewhat straightforward. Another peak candidate is found at around 3117 MeV/$c^2$, where the isobaric analogue state of S${}^0$(3115), which is denoted S${}^+$(3115) tentatively, is expected.

In order to obtain the mass and width of the S${}^+$(3140) by a function fitting procedure to the spectrum, we assume two Gaussians which individually represent the S${}^+$(3140) and S${}^+$(3115), and an exponential background. We fix the center and $\sigma$ of the second Gaussian to those values of S${}^0$(3115) obtained by a global fitting of the spectrum as shown in the top, assuming a third-order polynomial background and a Gaussian signal. It should be noted that the yield of S${}^+$(3115) is treated as a free parameter. In this functional fitting to the neutron higer-S/N spectrum, we obtain $\chi^2/DOF = 0.92$, the S${}^+$(3140) yield of $120.2 \pm 32.3$ counts, which results in a peak significance of 3.7 $\sigma$, a mass of $3140.5 \pm 2.3(stat)$ MeV/$c^2$, and standard deviation of $7.3 \pm 1.7(stat)$ MeV/$c^2$. The resulting functions are overlayed on the higher-S/N data points. The systematic uncertainty of the absolute mass is estimated to be ${}^{+3.0}_{-0.8}$ from the $1/\beta$ ambiguity due to the uncertainty of TOF origin, and the upper limit of the natural width is deduced to be 21.6 MeV/$c^2$ at 95 $\%$ confidence level, by using the lower limit of missing mass resolution of the measurement, 4.5 MeV/$c^2$($\sigma$). The main decay modes of the state are $\Sigma^+nn$ and/or $\Sigma^-pp$, according to the fact that the peak structure appears only if a fast secondary pion with $p_{\pi} > 125$ MeV/$c$ is detected in coincidence \cite{E471PLB_N}. The quantitative estimation of the formation probability is rather difficult, but it is expected to be similar to that of the S${}^0$(3115) in connection with the comparison of the yields between S${}^+$(3115) and S${}^0$(3115). 

\section{Isospin of the S${}^+$(3140) and consistency between the ${}^4$He($K^-_{stopped},\thinspace p$) and ${}^4$He($K^-_{stopped}, \thinspace n$) spectra}

 The total isospin of the S${}^+$(3140) is considered to be 0 as follows. Assumig S${}^+$(3140) having had $T=1$, we would obtain an equality,
\begin{equation}
 B_{{\rm S}^+(3115)} : B_{{\rm S}^0(3115)} = B_{{\rm S}^+(3140)} : B_{{\rm S}^0(3140)},
\end{equation}
which would be easily deduced from the isospin relation, where $B_{{\rm S}^{0/+}(3115/3140)}$ denotes the formation ratio per stopped $K^-$ of each strange tribaryon with $T=1$. This equality, or its equivalent expression, $B_{{\rm S}^+(3115)} : B_{{\rm S}^+(3140)} = B_{{\rm S}^0(3115)} : B_{{\rm S}^0(3140)}$, would be almost directly converted to the one between the experimental yields,
\begin{equation}
 Y_{{\rm S}^+(3115)} : Y_{{\rm S}^+(3140)} = Y_{{\rm S}^0(3115)} : Y_{{\rm S}^0(3140)}, 
\end{equation}
where $Y_{{\rm S}^{0/+}(3115/3140)}$ denotes the total yield of each state. This equality would claim that we should observe a peak structure at around 3140 MeV/$c^2$ on the proton spectrum with larger yield than that of S${}^0$(3115), but no such a significant structure has been observed, in fact. Therefore, we assign 0 to the total isospin of the S${}^0$(3140).

The total isospin of the S${}^0$(3115) is 1, while that of the S${}^+$(3140) is 0. Therefore, we do not have any simple relationship between the formation ratio of those two. If we look at the $T=1$ sector, it would seem strange at the first glance that S${}^0$(3115) has been observed with very large significance, while the indication of its isobalic analogue state, S${}^+$(3115), is not definitive. Now, we investigate this apparent contradiction. The formation ratios of the two $T=1$ analogue states populated by the reactions (\ref{eqn1}) and (\ref{eqn2}) are connected by a simple relationship which is deduced from the Clebsch-Gordan coefficients involving the two final states,
\begin{equation}
 Br_{{\rm S}^+(3115)} : Br_{{\rm S}^0(3115)} = (\sqrt{1/3})^2 : (-\sqrt{2/3})^2 = 1:2.
\end{equation}
 Now, the expected yields for both states in the experiment can be estimated by the relation,
\begin{equation}
\frac{Y_{{\rm S}^+(3115)}}{Y_{{\rm S}^0(3115)}}=\frac{Br_{{\rm S}^+(3115)}}{Br_{{\rm S}^0(3115)}} \frac{D_n}{D_p} \frac{Dc_{{\rm S}^+(3115)}}{Dc_{{\rm S}^0(3115)}},
\end{equation}
where $D_p$ and $D_n$ denote proton and neutron detection efficiency (neutron detection efficiency at 10 MeV electron equivalent) of NC array, respectively, and $Dc_{{\rm S}^+(3115)}$ and $Dc_{{\rm S}^0(3115)}$ are secondary charged particle detection probabilities by vertex detector system per primary decay, which are rather sensitive to the decay branching ratio of each state. The relative detection efficiency $D_n / D_p$ is known to be $\sim 0.25$ at around the energy region of interest, and we obtain
\begin{equation}
Y_{{\rm S}^+(3115)} : Y_{{\rm S}^0(3115)} \sim 1 : 8 
\end{equation}
assuming $Dc_{{\rm S}^0(3115)} \sim Dc_{{\rm S}^+(3115)}$. We can compare this approximate relation with the results of the functional fittings already presented in Fig. \ref{n_missingmass}. In those fittings, we have obtained $Y_{{\rm S}^0(3115)} = 428 \pm 88$ and $Y_{{\rm S}^+(3115)} = 73 \pm 25$ counts, respectively, which are fairly consistent with the equality. {\bf Therefore, the large yield of S${}^0$(3115) does not contradict the fact that its isobaric analogue state has not appeared clearly on the neutron spectrum}.

\section{Summary of the results and discussion}
 The properties of the observed strange tribaryons are tabulated below, together with the $\bar{K}$ binding energies with respect to $K^-+p+n+n$ and $K^-+p+p+n$ rest mass for S${}^0$(3115) and S${}^+$(3140), respectively. These results are compared to several theoretical calculations in the following.

\vspace{2mm}
\begin{tabular}{|c|c|c|}
\hline
\hline
state  & S${}^0$(3115) & S${}^+$(3140) \\
\hline
 reaction & ${}^4$He($K^-_{\rm stopped},p)$ & ${}^4$He($K^-_{\rm stopped},n)$ \\
\hline
 mass (MeV/$c^2$) & $3117.7^{+3.8}_{-2.0}(syst) \pm 0.9(stat)$  & $3140.5^{+3.0}_{-0.8}(syst) \pm 2.3(stat)$ \\
\hline
 $\bar{K}$ binding energy (MeV) & $193.4^{+2.0}_{-3.8}(syst) \pm 0.9(stat)$ & $169.3^{+0.8}_{-3.0}(syst) \pm 2.3(stat) $      \\
\hline
natural width (MeV/$c^2$) &  $< 21.6$ & $< 21.6$ \\
\hline
isospin $|T,T_z>$ & $|1,-1>$ & $|0,0>$($|1,0>?$) \\
\hline
charge            & 0        & +1              \\
\hline
deduced         &$\Sigma^0 nn$ &$\Sigma^+ nn$ and/or $\Sigma^- pp$ \\
decay modes     &$\Sigma^- np$ and/or $\Sigma^- d$ &  \\
\hline
formation ratio & below 1 $\%$   & below 1 $\%$ \\
\hline
\hline
\end{tabular}

\begin{wrapfigure}[20]{l}{10cm}
     \includegraphics[width=9cm]{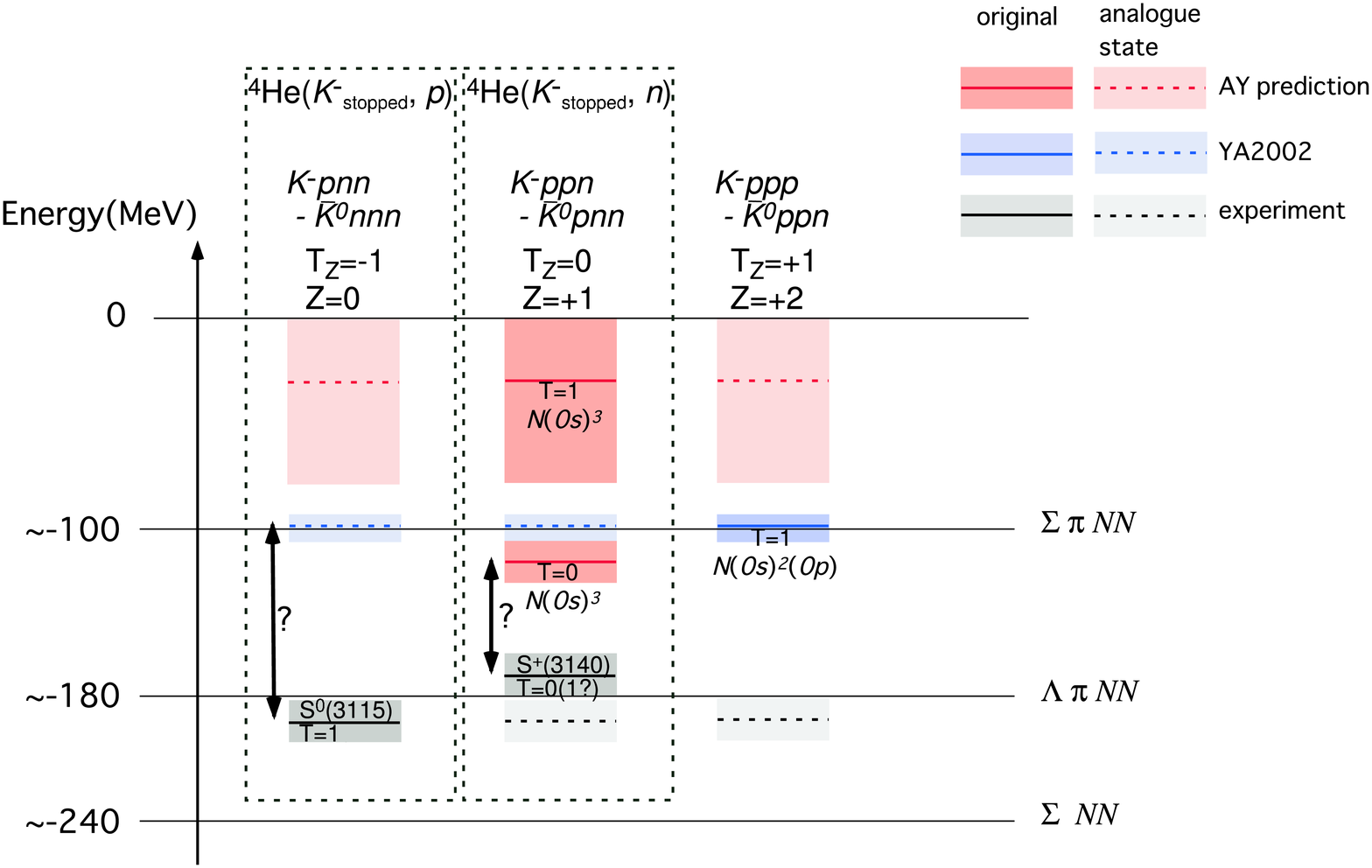}
  \caption{A comparison between non-relativistic calculations on the framework of deeply bound kaonic nuclear states and experimental results. The original\cite{Akaishi-Yamazaki}, subsequent\cite{YA2002,Dote} predictions and experimental results are  represented by red, blue and grey, respectively.}\label{ex-th-comp}
\end{wrapfigure}

\begin{description}
\item[Deeply bound $\bar{K}NNN$ state]{On the standpoint of deeply bound kaonic nuclear state, the discovery of the narrow and light S${}^0$(3115) state was quite unexpected, because $\bar{K}NNN^{Z=+1}_{|T,T_z> = |1,0>}$ with a triton-like nuclear core $[NNN]_{T=1/2}$ was expected to be very shalow and broad in their original prediction \cite{Akaishi-Yamazaki}. On the other hand, possible existence of a narrow deeply bound $\bar{K}N'NN^{Z=+2}_{|T,T_z>=|1,+1>}$ state with a $[N'NN]_{T=3/2}$ core in which $N'$ occupies the nuclear $0p$ orbit, has been predicted \cite{YA2002,Dote}.  Then, S${}^0$(3115) can be interpreted as the isobalic analogue state of predicted $\bar{K}N'NN^{Z=+2}_{|T,T_z>=|1,+1>}$ with $J^{\pi} = \frac{3}{2}^+$, $\bar{K}N'NN^{Z=0}_{|T,T_z>=|1,-1>}$, while S${}^+$(3140) can be identified as the originally predicted $\bar{K}NNN^{Z=+1}_{T=0}$ with $J^{\pi} = \frac{1}{2}^-$. A comparison between the calculated and observed binding energies is given in Fig. \ref{ex-th-comp}. In the comparison, we meet two difficulties, namely, the large difference of the binding energies up to $\sim 100$ MeV and the level reversing between the $T=0$ and $T=1$ states. However, it is pointed out that these difficulties can be resolved by taking care of the relativistic effect and by invoking an enhanced $\bar{K} N$ interaction and a strong spin-orbit interaction in the dense nuclear medium \cite{ADY}.}

\item[SU(3) multiskirmion]{On the framework of the skirmion description of multibaryon systems \cite{Skirmion}, $B=3$ and $S=-1$ systems have the excitation energies of 264 and 325 MeV for $(T,\thinspace J^{\pi})=(0,\thinspace \frac{1}{2}^+)$ and $(0,\thinspace \frac{5}{2}^-)$, respectively, while it gives 304 MeV for the $(1,\thinspace \frac{1}{2}^+)$ system. These excitation energies, which are defined with respect to the $S=0$ lowest energy state, correspond to the $\bar{K}$ binding energies of 230, 169 and 190 MeV, respectively.}

\item[Nona-quark]{ Very recently, a nona-quark interpretation of the strange tribaryons is suggested \cite{Nona_quark}. In this framework, S${}^0$(3115) is identified as a member of flavor 27-plet with $(T,\thinspace J) = (1, \frac{1}{2})$ or $(1, \frac{3}{2})$, and observed narrow width may be explained by the small overlap of the wave functions of S${}^0$ and hadronic final states. S${}^+$(3140) is then identified as the state with $(F,I,J)=(10^*,0,3/2)$ or $(35^*,0,1/2)$.
}
\end{description}

There are many differences among these theories with respect to the quantum numbers, and thus, the experimental determination of spin-parity of these strange stribaryons are awaited.

In conclusion, we have observed two kinds of strange tribaryons in the ${}^4$He($K^-_{stopped},\thinspace N$) reactions. One species, S${}^0$(3115), which was observed in the proton emission channel, has $M_{{\rm S}^0} \sim 3118$ MeV/$c^2$ and isospin 1. The other species, S${}^+$(3140) indicated from the neutron emission channel, has $M_{{\rm S}^+} \sim 3140$ MeV/$c^2$ and its isospin is assigned to be 0. Several possible interpretations exist, but we cannot discriminate them at the moment.

\end{document}